\documentclass[12pt]{article} 
\usepackage[sectionbib]{natbib}
\usepackage{array,epsfig,fancyheadings,rotating}
\usepackage[]{hyperref}  
\usepackage{sectsty, secdot}
\sectionfont{\fontsize{12}{14pt plus.8pt minus .6pt}\selectfont}
\renewcommand{\theequation}{\thesection\arabic{equation}}
\subsectionfont{\fontsize{12}{14pt plus.8pt minus .6pt}\selectfont}

\usepackage{amsmath}
\usepackage{amssymb}
\usepackage{amsfonts}
\usepackage{multirow}
\usepackage{amsthm}

\usepackage{natbib}

\usepackage{indentfirst}
\usepackage{graphicx}
\usepackage{subcaption}
\usepackage{amsmath}
\usepackage{amssymb}
\usepackage{bm}
\usepackage{multirow}
\usepackage{booktabs}
\usepackage{siunitx}
\usepackage{makecell}
\usepackage{float}
\usepackage{titlesec}
\usepackage{diagbox}
\usepackage{hyperref}
\usepackage[T1]{fontenc}
\usepackage{currvita}
\usepackage[applemac]{inputenc}
\usepackage{tabu}
\usepackage{times}
\usepackage{algorithm, algpseudocode}
\usepackage{amssymb, amsmath, mathrsfs, amsfonts, graphicx, multirow, url}
\usepackage{adjustbox}
\usepackage{array, epsfig, rotating, pdflscape, bm}
\usepackage{algorithm, algpseudocode}
\usepackage{enumerate}
\usepackage{etoolbox}
\usepackage{xcolor}

\usepackage{tikz}
\usepackage{pict2e,picture}
\usetikzlibrary{shapes.symbols, shapes.geometric, arrows, positioning, chains}
\numberwithin{equation}{section}

\def\bSig\mathbf{\Sigma}

\newcommand{\bc} {\mathbf {c}}

\newcommand{\be} {\mathbf {e}}

\newcommand{\bI} {\mathbf {I}}

\newcommand{\bP} {\mathbf {P}}

\newcommand{\bq} {\mathbf {q}}

\newcommand{\br} {\mathbf {r}}

\newcommand{\bfs} {\mathbf{s}}

\newcommand{\bt} {\mathbf {t}}

\newcommand{\bu} {\mathbf {u}}
\newcommand{\bU} {\mathbf {U}}

\newcommand{\bv} {\mathbf {v}}

\newcommand{\bx} {\mathbf {x}}

\newcommand{\bz} {\mathbf {z}}

\newcommand{\calA}{\mathcal{A}}
\newcommand{\calB}{\mathcal{B}}
\newcommand{\calN}{\mathcal{N}}
\newcommand{\calT}{\mathcal{T}}

\newcommand{\bbR} {\mathbb{R}}

\newcommand{\inprob}{\stackrel{p}{\rightarrow}}
\newcommand{\indist}{\stackrel{d}{\rightarrow}}

\DeclareMathOperator*{\argmin}{arg\,min}
\DeclareMathOperator*{\argmax}{arg\,max}

\newcommand{\bbeta} {\mbox{\boldmath $\beta$}}
\newcommand{\bdelta} {\mbox{\boldmath $\delta$}}

\newcommand{\bxi} {\mbox{\boldmath $\xi$}}

\newcommand{\sign}{\operatorname{sign}}
\newcommand{\calC}{\mathcal{C}}

\newcommand{\calM}{\mathcal{M}}
\newcommand{\calQ}{\mathcal{Q}}

\newtheorem{theorem}{Theorem}
\newtheorem{assumption}{Assumption}

\newtheorem{corollary}{Corollary}

\theoremstyle{definition}

\begin{document}


\renewcommand{\baselinestretch}{2}

\markright{ \hbox{\footnotesize\rm Statistica Sinica
}\hfill\\[-13pt]
\hbox{\footnotesize\rm
}\hfill }

\markboth{\hfill{\footnotesize\rm FIRSTNAME1 LASTNAME1 AND FIRSTNAME2 LASTNAME2} \hfill}
{\hfill {\footnotesize\rm Post-selection inference in GLMs} \hfill}

\renewcommand{\thefootnote}{}
$\ $\par


\fontsize{12}{14pt plus.8pt minus .6pt}\selectfont \vspace{0.8pc}
\centerline{\large\bf Post-selection inference in generalized linear models}
\vspace{2pt} 
\centerline{\large\bf via parametric programming}
\vspace{.4cm} 
\centerline{Qinyan Shen, Karl Gregory, Xianzheng Huang} 
\vspace{.4cm} 
\centerline{\it University of South Carolina}
 \vspace{.55cm} \fontsize{9}{11.5pt plus.8pt minus.6pt}\selectfont


\begin{quotation}
\noindent {\it Abstract:}
We propose a unified framework to draw inferences for regression coefficients in a generalized linear model (GLM) following Lasso-based variable selection. We adapt to non-Gaussian GLMs a recently developed parametric programming strategy for post-selection inference in the linear model with a Gaussian response by drawing parallels between maximum likelihood estimation in GLMs and least squares estimation in linear models. We then conduct post-selection inference based on a linearized model for pseudo response and covariate data strategically created based on the raw data. Using synthetic data generated from regression models for three different types of non-Gaussian responses in simulation experiments, we demonstrate that the proposed method effectively corrects the naive inference that ignores variable selection while achieving greater efficiency than a polyhedral-based post-selection adjustment.

\vspace{9pt}
\par
\end{quotation}\par

\def\thefigure{\arabic{figure}}
\def\thetable{\arabic{table}}

\renewcommand{\theequation}{\thesection.\arabic{equation}}

\fontsize{12}{14pt plus.8pt minus .6pt}\selectfont

\section{Introduction}

Traditional statistical inference assumes all hypotheses of interest are formulated prior to the observation of data.   In regression contexts, practitioners often explore their data in order to select from a set of available variables a subset to use as covariates and, after fitting a model with these covariates, wish to make inferences on their effects. Such a strategy permits the data to dictate which hypotheses are ultimately tested, wherein lies the danger of incurring higher Type I error rates than intended. This danger has motivated the development of many post-selection inference methods to account for model selection.  We may categorize these as \textit{data splitting methods} \citep{wasserman2009high, meinshausen2009p, rinaldo2019bootstrapping, rasines2023splitting}, \textit{simultaneous inference methods} \citep{berk2013valid, zhang2017simultaneous, bachoc2019valid, bachoc2020uniformly}, and \textit{conditioning methods} \citep{lee2016exact, tibshirani2016exact, taylor2018post, kuchibhotla2020valid, pirenne2024parametric, neufeld2022tree}. 

We here pursue a conditioning method, whereby we make inferences on a parameter in the selected model based on the conditional sampling distribution of a relevant statistic given the selection event. The seminal work \citet{lee2016exact} studied the sampling distribution of a linear contrast of Gaussian responses in a linear model given selection via $L_1$-penalization of the least-squares criterion, showing that the selection event can be characterized as the union of many polyhedra in the support of the response data. To obtain a more tractable sampling distribution of the statistic, these authors further condition on the signs of the selected regression coefficients. This additional conditioning costs efficiency in statistical inference, leading to wider-than-necessary confidence intervals that, although guaranteeing a nominal coverage probability, could have infinite expected width \citep{kivaranovic2021length}. Some recent developments in this vein extend beyond linear models and Lasso regularization, where one seeks a useful conditional sampling distribution of the target statistic after intersecting the selection event with additional characteristics of the selected model so that the conditional event is (approximately) polyhedral in the support of the response data \citep{panigrahi2023approximate, zhao2022selective, shen2024post,taylor2018post}. We refer to these methods as polyhedral methods.

\citet{le2021parametric} made improvements to the work of \citet{lee2016exact} by introducing a parametric programming (PP) approach to find the sampling distribution of a linear contrast of the response data conditional only on the selection event, avoiding the efficiency loss incurred by the sign-conditioning of the polyhedral method. \citet{pirenne2024parametric} extended the PP approach to inference following  model selection via adaptive Lasso, adaptive elastic net, and group Lasso. In our work we adapt the PP approach to generalized linear models (GLMs) for non-Gaussian responses.

Our proposed strategy consists of two steps. Section~\ref{sec:step1} provides the development of the first step, in which we ``linearize'' the regression problem specified by a GLM for non-Gaussian data. Section~\ref{sec:pp} elaborates on the second step, where we apply the PP method to the linearized regression model.  Section~\ref{sec:3models} describes in detail the implementation of the proposed method in three non-Gaussian regression settings. Section ~\ref{sec:sim} presents simulation studies comparing the proposed method with the naive method (the method which ignores model selection), the polyhedral method. Section~\ref{sec:real} presents three case studies in which we make inferences on covariate effects following variable selection using different types of non-Gaussian data arising from real-life applications. Section~\ref{sec:disc} outlines key takeaways and suggestions for future research.

\section{Pre-selection inference in GLMs}
\label{sec:step1}
Suppose we observe $(\bx_1,Y_1),\dots,(\bx_n,Y_n)$, where $\bx_1,\dots,\bx_n \in \mathbb{R}^p$ are fixed covariate vectors and $Y_1,\dots,Y_n$ are independent responses such that
\begin{equation}
\label{eqn:glm}
Y_i \sim f(y;\eta_i,\phi) = \exp \left\{\frac{y\eta_i - b(\eta_i)}{a(\phi)} - c(y,\phi)\right\},
\end{equation}
where $\eta_i \equiv  \beta_0 + \bx_i^\top\bbeta$, for $i = 1,\dots,n$, where $a(\cdot)$, $b(\cdot)$ and $c(\cdot,\cdot)$ are known functions, $\phi$ is a dispersion parameter, and $\beta_0$ and $\bbeta \equiv (\beta_1 ,\dots, \beta_p)^\top$ are parameters with unknown values. This is the canonical generalized linear model (GLM); see \cite{mccullagh1989generalized}. Note that the mean and variance of $Y_i$ are $b'(\eta_i)$ and $a(\phi)b''(\eta_i)$, respectively, where $b'(\cdot)$ and $b''(\cdot)$ are the first two derivatives of $b(\cdot)$. Throughout, let $M_0 = \{j \in \{1,\dots,p\}: \beta_j \neq 0\}$ be the set of indices corresponding to nonzero regression coefficients. We will consider making inferences following the selection of a model $M \subset \{1,\dots,p\}$.

\subsection{Maximum likelihood estimation in a submodel}
\label{sec:mleglm}
To focus on making inferences on the regression coefficients in the GLM specified by \eqref{eqn:glm}, we will for most of the paper assume $\phi$ is known.  The maximum likelihood estimator (MLE), which we denote by $(\hat \beta_0,\hat \bbeta)$, of the tuple $(\beta_0,\bbeta)$ can be obtained by maximizing the log-likelihood
\[
\ell_\phi(t_0,\bt) \equiv \frac{1}{a(\phi)}\sum_{i=1}^n(Y_i\eta_i(t_0,\bt) - b(\eta_i(t_0,\bt))) - \sum_{i=1}^n c(Y_i,\phi)
\]
over $(t_0,\bt) \in \bbR\times \bbR^p$, where $\eta_i(t_0,\bt) \equiv t_0 + \bx_i^\top \bt$.  Instead of including all covariates, one may consider selecting a model $M\subset \{1, \ldots, p\}$ indexing which covariates to include in the construction of the linear predictors $\eta_1,\dots,\eta_n$.

The MLE in model $M$, denoted by $(\hat \beta_{0,M}^\text{mle}, \hat \bbeta{}^\text{mle}_{M})$, maximizes the function
\[
\ell_{\phi,M}(t_0,\bt) \equiv \frac{1}{a(\phi)}\sum_{i=1}^n(Y_i\eta_{i,M}(t_0,\bt) - b(\eta_{i,M}(t_0,\bt))) - \sum_{i=1}^n c(Y_i,\phi)
\]
over $(t_0,\bt) \in \bbR\times \bbR^{|M|}$, where $\eta_{i,M}(t_0,\bt) \equiv t_0 + (\bx_i)_M^\top\bt$, $|M|$ is the cardinality of $M$, and $(\bx_i)_M$ is the vector constructed from the entries of $\bx_i$ with indices in $M$. Under regularity conditions \citep{white1982maximum}, as $n\to \infty$,  
 $(\hat \beta{}^\text{mle}_{0,M}, \hat \bbeta{}^\text{mle}_{M})$ converges in probability to the limiting value of 
\[
(\beta_{0,M}^*,\bbeta_{M}^*) \equiv \argmax_{ (t_0, \bt) \in \bbR\times \bbR^{|M|}} E \frac{1}{n}\{\ell_\phi(\beta_0,\bbeta) - \ell_{\phi,M}(t_0, \bt)\},
\]
where the maximand is proportional to the Kullback-Leibler divergence of the true model from model $M$. One may therefore regard $(\beta_{0,M}^*,\bbeta_{M}^*)$ as the target of estimation when model $M$ is considered. Denoting by $\bbeta_M$ the vector containing the entries of $\bbeta$ with indices in $M$, we will have $(\beta_{0,M}^*,\bbeta_{M}^*)=(\beta_0,\bbeta_{M})$ if $M$ contains all the indices in which $\bbeta$ is nonzero, that is if $M \supset M_0$.

If model $M$ results from variable selection based on the observed data, valid inferences for $\bbeta_{M}^*$ using the same data should account for the selection event. In this study, we focus on Lasso-based variable selection. For example, one may  select a model by finding the $L_1$-penalized MLE, defined as
\begin{equation}
\label{eqn:lassoMLE}
(\hat \beta{}^\text{mle}_{0,\lambda},\hat \bbeta{}^\text{mle}_\lambda) \equiv \argmin_{(t_0,\bt) \in  \bbR \times \bbR^p} \Big\{ - \frac{1}{n}\ell_\phi(t_0,\bt) + \lambda \|\bt\|_1 \Big\},
\end{equation}
where $\|\bt\|_1$ is the $L_1$-norm of $\bt$ and $\lambda > 0$ is a tuning parameter governing the sparsity of $\hat \bbeta{}^\text{mle}_\lambda$. A selected model is then $\hat M_\lambda^{\text{mle}} \equiv \{j :  (\hat{\bbeta}{}^{\text{mle}}_\lambda)_j \neq 0\}$. To make inferences on  $(\bbeta_M^*)_j$ based  $(\hat \bbeta{}^{\text{mle}}_M)_j$, where for a generic vector $\bv$ we denote by $(\bv)_i$ entry $i$ of $\bv$, it is necessary to obtain the conditional sampling distribution of $(\hat \bbeta{}^{\text{mle}}_M)_j$ given the selection event $\{\hat M{}^\text{mle}_\lambda = M\}$. However, this conditional distribution does not appear to be analytically tractable in the GLM setting.  For this reason we pursue a two-step strategy whereby we first ``linearize'' the GLM and then apply a post-selection inference method developed for the linear model.

\subsection{Model linearization}
\label{sec:glmlinear}

In linear regression with a Gaussian response, the exact conditional distribution of $(\hat \bbeta{}^\text{mle}_{M})_j | \{\hat M_\lambda^\text{mle} = M\}$ has been found \citep{lee2016exact, le2021parametric}. This motivates our strategy of linearizing the GLM and then performing model selection and post-selection inference in the linearized model. Our linearization step is inspired by the Newton-Raphson (NR) update leading to the MLEs of $(\beta_0,\bbeta)$, which we describe next.

For each $(t_0,\bt) \in \bbR \times \bbR^p$ define $\bz(t_0,\bt)$  as the vector with entries
\[
z_i(t_0,\bt) \equiv \sqrt{b''(\eta_i(t_0,\bt))}\eta_i(t_0,\bt) + \frac{Y_i - b'(\eta_i(t_0,\bt))}{\sqrt{b''(\eta_i(t_0,\bt))}},
\]
the vector $\bu_0(t_0,\bt)$ with entries $\sqrt{b''(\eta_i(t_0,\bt))}$, and the matrix $\bU(t_0,\bt)$ with rows $\sqrt{b''(\eta_i(t_0,\bt))}\bx_i^\top$ for $i=1,\dots,n$. Then, with initial value $(t_0^{(0)},\bt^{(0)})$, the iteratively reweighted least squares formulation of the NR update is
\begin{equation}
\label{eqn:nrls}
(t_0^{(k)} , \bt^{(k)}) \leftarrow \underset{(t_0,\bt) \in \bbR\times\bbR^p}{\argmin} \|\bz^{(k-1)} - (\bu_0^{(k-1)} t_0 + \bU^{(k-1)} \bt)\|^2,
\end{equation}
where $\bz^{(k-1)}  \equiv \bz(t_0^{(k-1)},\bt^{(k-1)})$, $\bu_0^{(k-1)} \equiv \bu_0(t_0^{(k-1)},\bt^{(k-1)})$, and $\bU^{(k-1)} \equiv \bU(t_0^{(k-1)},\bt^{(k-1)})$; see \cite{davison2003statistical}.

Upon convergence of $(t_0^{(k)} , \bt^{(k)})$ to $(\hat \beta_0,\hat \bbeta)$, define $\hat \bz \equiv \bz(\hat \beta_0,\hat \bbeta)$, $\hat \bu_0 \equiv \bu(\hat \beta_0,\hat \bbeta)$, and $\hat \bU \equiv \bU(\hat \beta_0,\hat \bbeta)$. To focus on the parameters in $\bbeta$, we now define ``centered'' versions of the response vector $\hat \bz$ and the design matrix $\hat \bU$. First define $\bP_0(t_0,\bt) \equiv \|\bu_0(t_0,\bt)\|^{-2}\bu_0(t_0,\bt)\bu_0(t_0,\bt)^\top$ as well as $\bz_0(t_0,\bt) = (\bI - \bP_0(t_0,\bt))\bz(t_0,\bt)$ and $\bU_0(t_0,\bt) \equiv (\bI - \bP_0(t_0,\bt))\bU(t_0,\bt)$. Then set $\hat \bP_0 \equiv \bP_0(\hat \beta_0,\hat \bbeta)$ as well as $\hat \bz_0 \equiv (\bI - \hat \bP_0)\hat \bz$ and $\hat \bU_0 \equiv (\bI - \hat \bP_0)\hat \bU$, noting that $\hat\bP_0$ is the orthogonal projection onto the space spanned by the ``intercept'' vector $\hat \bu_0$.

Then we have $\hat\bbeta = \argmin_{\bt \in \bbR^p} ~ \|\hat \bz_0 - \hat \bU_0 \bt\|^2$, so that  we may regard $\hat \bbeta$ as the least squares estimator in linear regression with response vector $\hat \bz_0$ and  design matrix $\hat \bU_0$. From here our strategy will be to treat the response vector $\hat \bz_0$ as though it arose from a Gaussian linear model with design matrix $\hat \bU_0$ and to apply a post-selection inference method developed for Gaussian linear regression.

Since our strategy is to treat $\hat \bz_0$ and $\hat \bU_0$ as Gaussian linear model data, we propose to select a model via $L_1$-penalization of the Gaussian log-likelihood with $\hat \bz_0$ and $\hat \bU_0$ plugged in. That is, we propose computing the sparse estimator
\begin{equation}
\label{eqn:betalambda}
\hat\bbeta_\lambda \equiv \underset{\bt \in \bbR^p}{\argmin} ~ \frac{1}{2n}\|\hat \bz_0 - \hat \bU_0 \bt\|^2 + \lambda \|\bt\|_1,
\end{equation}
for some $\lambda > 0$ and selecting the model $\hat M_\lambda \equiv \{j : (\hat \bbeta_{\lambda})_j \neq 0\}$. Note that the model selected in this way may be distinct from the model $\hat M_\lambda^{\text{mle}}$ selected via $L_1$-penalization of the original GLM log-likelihood; however, in Section \ref{sec:ideal} we argue that these models should be reliably similar as $n \to \infty$.  Then, on the event that the model $\hat M_\lambda = M$ is chosen, we consider the conditional distribution of the vector $\hat\bbeta_M \equiv \underset{\bt \in \bbR^p}{\argmin} ~\|\hat \bz_0 - \hat \bU_{0,M} \bt\|^2$, the entries of which may be expressed as contrasts in the vector $\hat \bz_0$ of the form
\begin{equation}
\label{eqn:conthat}
(\hat \bbeta_M)_j = \be_j^T(\hat \bU_{0,M}^\top\hat \bU_{0,M})^{-1}\hat \bU_{0,M}^\top\hat \bz_0,
\end{equation}
for $j =1,\dots,|M|$, where $\be_j$ is the $|M|\times 1$ vector with entry $j$ equal to one and remaining entries equal to zero.

In order to study the conditional distributions of contrasts of the form in \eqref{eqn:conthat}, we next introduce idealized counterparts to $\hat \bz_0$ and $\hat \bU_0$, which we denote by $\bz_0$ and $\bU_0$, which one would observe if one knew the true values of the parameters $\beta_0$ and $\bbeta$.

\subsection{The idealized linear model}
\label{sec:ideal}

The response vector $\hat \bz$, the vector $\hat \bu_0$, and design matrix $\hat \bU$ can be viewed as approximations to unobservable, idealized counterparts $\bz$, $\bu_0$, and $\bU$, respectively, which we define as $\bz \equiv \bz(\beta_0,\bbeta)$, $\bu_0 \equiv \bu_0(\beta_0,\bbeta)$, and $\bU\equiv \bU(\beta_0,\bbeta)$. Moreover, letting $\bz_0 \equiv \bz_0(\beta_0,\bbeta)$ and $\bU_0 \equiv \bU_0(\beta_0,\bbeta)$, we obtain idealized counterparts to the ``centered'' response vector $\hat \bz_0$ and design matrix $\hat \bU_0$.

If one could observe $\bz_0$ and $\bU_0$, one could base inferences on the idealized linear estimator of $\bbeta$ given by $\tilde \bbeta \equiv \argmin_{\bt \in \bbR^p} \|\bz_0 - \bU_0 \bt\|^2$. Likewise, one could compute the idealized sparse estimator 
\begin{equation}
\label{eqn:betalambdaideal}
\tilde \bbeta_\lambda \equiv \underset{\bt \in \bbR^p}{\argmin} ~\frac{1}{2n}\|\bz_0 -\bU_0 \bt\|^2 + \lambda \|\bt\|_1,
\end{equation}
and the corresponding idealized selected model $\tilde M_\lambda \equiv \{j : (\tilde \bbeta_\lambda )_j \neq 0\}$. Furthermore, on the event $\tilde M_\lambda = M$, one could make conditional inferences by considering the conditional distributions of the entries of the idealized estimator $\tilde \bbeta{}_M \equiv \underset{\bt \in \bbR^p}{\argmin} \|\bz_0 - \bU_{0,M} \bt\|^2$, which could be expressed as contrasts in the vector $\bz_0$ with form
\begin{equation}
\label{eqn:cont}
(\tilde  \bbeta{}_M)_j \equiv \be_j^\top( \bU_{0,M}^\top \bU_{0,M})^{-1} \bU_{0,M}^\top \bz_0.
\end{equation}
Note that the contrast in \eqref{eqn:cont} is an idealized version of the contrast in \eqref{eqn:conthat}.  Defining for each $M \subset \{1,\dots,p\}$ and $j = 1,\dots,|M|$ the vector
\[
\bc_{M,j}(t_0,\bt)^\top = \be_j^\top( \bU_{0,M}(t_0,\bt)^\top \bU_{0,M}(t_0,\bt))^{-1} \bU_{0,M}(t_0,\bt)^\top,
\]
where $\bU_{0,M}(t_0,\bt)$ is the matrix formed with the columns of $\bU_{0}(t_0,\bt)$ having indices in $M$, we set $\hat \bc_{M,j} \equiv \bc_{M,j}(\hat \beta_0,\hat \bbeta)$, which has idealized counterpart  $\bc_{M,j} \equiv \bc_{M,j}(\beta_0,\bbeta)$. This allows us to write \eqref{eqn:conthat} and \eqref{eqn:cont} as $(\hat \bbeta{}_M)_j =  \hat \bc_{M,j}^\top\hat\bz_0$ and $(\tilde \bbeta{}_M)_j =  \bc_{M,j}^\top\bz_0$, respectively.


Our first result gives conditions under which, prior to model selection, one can make inferences based on the observable $\hat \bz_0$ and $\hat \bU_0$ which are asymptotically equivalent to those based on their idealized counterparts $\bz_0$ and $\bU_0$.  To state our result, define for any $(t_0,\bt) \in \bbR\times \bbR^p$ the function
\[
g_{M,j,n}(t_0,\bt;\bbeta) \equiv \frac{\bc_{M,j}(t_0,\bt)^\top(\bz(t_0,\bt)- \bU(t_0,\bt)\bbeta)}{\sqrt{\bc_{M,j}(t_0,\bt)^\top\bc_{M,j}(t_0,\bt)}}.
\]
Then the function $g_{M,j,n}(t_0,\bt;\bbeta)$ may be used to construct some useful pivotal quantities. Specifically we define
\[
\hat g_{M,j,n}(\bbeta) \equiv g_{M,j,n}(\hat \beta_0,\hat \bbeta ; \bbeta) =\frac{\hat \bc_{M,j}^\top(\hat \bz_0 - \hat \bU_0\bbeta)}{\sqrt{\hat \bc_{M,j}^\top \hat \bc_{M,j}}}
\]
as a feasible pivotal quantity and 
\[
g_{M,j,n}(\bbeta) \equiv g_{M,j,n}( \beta_0,\bbeta ; \bbeta)  = \frac{\bc_{M,j}^\top( \bz_0 -  \bU_0\bbeta)}{\sqrt{\bc_{M,j}^\top \bc_{M,j}}}
\]
as its idealized counterpart.  After stating an assumption, we can present our first main result.

\begin{assumption}
\label{assum:main}
Given $M \subset \{1,\dots,p\}$ and an index $j =1,\dots,|M|$, suppose (i) $\|\bc_{M,j}\|_\infty / \|\bc_{M,j}\|\to 0$ as $n \to\infty$ and (ii) for some $n_0 \geq 1$ and $\delta, C \in [0,\infty)$,
\[
\mathbb{E} |g_{M,j,n}(t_0,\bt;\bbeta) - g_{M,j,n}(\beta_0,\bbeta;\bbeta)| \leq C \|(t_0,\bt^\top)^\top - (\beta_0,\bbeta{}^\top)^\top\|
\]
for all $n > n_0$ for all $(t_0,\bt)$ such that $\|(t_0,\bt^\top)^\top - (\beta_0,\bbeta{}^\top)^\top\| \leq \delta$.
\end{assumption}

Assumption 1(i) is mild and holds if the maximum leverage in the linear model with design matrix $\bU_M$ converges to zero; see \cite{huber2011robust}. Assumption 1(ii) is a smoothness condition on the function $g_n(t_0,\bt;\bbeta)$ in the neighborhood of the tuple $(\beta_0,\bbeta)$. Namely, it requires that the expected change in the (random) function $g_{M,j,n}(t_0,\bt;\bbeta)$ as $(t_0,\bt)$ moves away from $(\beta_0,\bbeta)$ is bounded above by some constant times the distance between $(t_0,\bt)$ and $(\beta_0,\bbeta)$. Both assumptions describe conditions on the sequence of design vectors $\{\bx_n\}_{n\geq 1}$.

\begin{theorem}
\label{thm:pivot}
Under Assumption \ref{assum:main} we have (i) $g_{M,j,n}(\bbeta) \indist  \calN (0,a(\phi))$ and (ii) $|\hat g_{M,j,n}(\bbeta) - g_{M,j,n}(\bbeta)| \inprob 0$ as $n \to \infty$.
\end{theorem}

The following corollary shows how Theorem \ref{thm:pivot} would enable inference on an entry of $\bbeta_M$ based on the observable $\hat \bz_0$ and $\hat \bU_0$.

\begin{corollary}
\label{cor:betajpivot}
Under Assumption \ref{assum:main}, if $M \supset M_0$ then, as $ n \to \infty$, we have
\[
\frac{(\hat \bbeta_{M})_j - (\bbeta_M)_j}{\sqrt{\be_j^T (\hat \bU_{0,M}^\top\hat \bU_{0,M})^{-1}\be_j}} \indist \calN(0,a(\phi)).
\]
\end{corollary}

Note that if $M \not\supset M_0$, then the estimation targets in the linearized model become the entries of the vector
\begin{equation}
\label{eqn:betastarm}
\tilde \bbeta{}_{M}^* \equiv ( \bU_{0,M}^\top\bU_{0,M})^{-1} \bU_{0,M}^\top\bU_0 \bbeta.
\end{equation}


Note that Theorem \ref{thm:pivot} and Corollary \ref{cor:betajpivot} give asymptotic distributions which are not yet conditioned on the selection of a model. In order to establish asymptotic equivalence of conditional inferences after model selection based on the observable linear model data $\hat \bz_0$ and $\hat \bU_0$ and those based on the idealized linear model data $\bz_0$ and $\bU_0$, we must investigate  whether the selected model $\hat M_\lambda$ based on the sparse estimator $\hat \bbeta_\lambda$ in \eqref{eqn:betalambda} and the selected model $\tilde M{}_\lambda$ based on $\tilde \bbeta_\lambda$ in \eqref{eqn:betalambdaideal} will agree with high probability.  If so, one may assume in every step of the analysis that one has observed the idealized response $\bz_0$ and  design matrix $\bU_0$ instead of their observable counterparts $\hat \bz_0$  and $\hat \bU_0$. Proofs of Theorem \ref{thm:pivot} and Corollary \ref{cor:betajpivot} are given in the Supplementary Material.

Defining the vector of correlations $\tilde \br_\lambda \equiv \tilde \bU_0^\top(\tilde \bz_0 - \tilde \bU_0\tilde \bbeta_\lambda)$, the KKT conditions give that $|(\tilde \br_\lambda)_j| = \lambda$ for all $j \in \tilde M_\lambda$ and $|(\tilde \br_\lambda)_j| \leq \lambda$ for all $j \notin \tilde M_\lambda$. Likewise defining $\hat \br_\lambda \equiv \hat \bU_0^\top(\hat \bz_0 - \hat \bU_0\hat \bbeta_\lambda)$, we have $|(\hat \br_\lambda)_j| = \lambda$ for all $j \in \hat M_\lambda$ and $|(\hat \br_\lambda)_j| \leq \lambda$ for all $j \notin \hat M_\lambda$.  If one assumes for the idealized pseudo-data that (i) $|(\tilde \br_\lambda)_j| \leq \lambda(1-\rho)$ for all $j \notin \tilde M_\lambda$ for some $\rho \in (0,1)$, a condition called strict dual feasibility \citep{wainwright2009sharp}, and (ii) $\min_{j \in \tilde M_\lambda}|(\tilde \bbeta_\lambda)_j| \geq c > 0$, a so-called beta-min condition \citep{zhao2006model}, hold with probability tending to one as $n \to \infty$, then $P(\hat M_\lambda = \tilde M_\lambda)\to 1$ provided
\begin{equation}
\label{eqn:rbconverge}
\|\hat \br_\lambda - \tilde \br_\lambda\| \inprob 0 \quad \text{ and }\quad  \|\hat \bbeta_\lambda - \tilde \bbeta_\lambda\| \inprob 0
\end{equation}
as $n \to \infty$. This is due to the fact that under (i), the first convergence in \eqref{eqn:rbconverge} implies that for all indices $j$ for which $|(\tilde \br_\lambda)_j| < \lambda$ (and therefore $(\tilde \bbeta_\lambda)_j = 0$), we will also have $|(\hat \br_\lambda)_j| < \lambda$ (and therefore $(\hat \bbeta_\lambda)_j = 0$). So we will have $(\hat \bbeta_\lambda)_j = 0$ for all $j$ such that $(\tilde \bbeta_\lambda)_j = 0$.  Under the beta-min condition (ii), the second convergence in \eqref{eqn:rbconverge} implies that for all $j$ such that $(\tilde \bbeta_\lambda)_j \neq 0$, we will have $(\hat \bbeta_\lambda)_j \neq 0$. 
The convergences in \eqref{eqn:rbconverge} can be established under mild smoothness and convexity conditions on the objective function defined by $q_n(\bfs;t_0,\bt) \equiv (2n)^{-1}\|\bz_0(t_0,\bt) - \bU_0(t_0,\bt)\bfs\|^2 + \lambda\|\bfs\|_1$, for which we can write $\tilde \bbeta_\lambda = \argmin_{\bfs}q_n(\bfs;\beta_0,\bbeta)$ and $\hat \bbeta_\lambda = \argmin_{\bfs}q_n(\bfs;\hat \beta_0,\hat \bbeta)$. Under such conditions, $\|(\hat \beta_0,\hat \bbeta{}^\top)^\top - (\beta_0,\bbeta{}^\top)^\top\|\inprob 0$ will imply the convergences in \eqref{eqn:rbconverge}. 

We next describe post-selection inference based on treating $\hat \bz_0$ and $\hat \bU_0$ as though they were equal to $\bz_0$ and $\bU_0$ and treating $\hat M_\lambda$ as though it matched $\tilde M_\lambda$.

\subsection{Variable selection after linearization}
\label{sec:glmvs}

Here we consider whether variable selection in the linear model with pseudo-data $\hat \bz_0$ and $\hat \bU_0$ will be perform similarly to variable selection based on $L_1$-penalization of the original GLM likelihood; that is, we consider how likely it is that selected models $\hat M_\lambda$ and $\hat M_\lambda^{\text{mle}}$ will match.  

We find that if the conditions are met for model selection consistency by $L_1$-penalization of the GLM likelihood, then the conditions are also met for model selection consistency by $L_1$-penalization in the linear model with the idealized pseudo-data $\bz_0$ and $\bU_0$. In particular, \cite{lee2015modelselection} present two main assumptions allowing for model selection consistency of a class of regularized M-estimators. These are a restricted strong convexity (RSC) condition and the so-called irrepresentable condition, the first version of which appeared in \cite{zhao2006model}. To express these conditions in our GLM setting with $L_1$-penalization, denote the Hessian of the scaled negative log-likelihood appearing in \eqref{eqn:lassoMLE} by $\calQ(t_0,\bt) \equiv (na(\phi))^{-1}\sum_{i=1}^n b''(\eta_i(t_0,\bt))\tilde \bx_i\tilde \bx_i^\top$, where $\tilde \bx_i \equiv [1 ~ \bx_i^\top]^\top$, $i=1,\dots,n$, and set $\calQ_\phi  \equiv \calQ(\beta_0,\bbeta)$. Now let $\calC_0 \times \calC \subset \bbR\times \bbR^p$ be a known convex set containing $(\beta_0,\bbeta)$ and set $\calM_0 \equiv \operatorname{span}\{\be_j,j \in M_0\}$, where $\{\be_j,j=1,\dots,p\}$ are elementary basis vectors in $\bbR^p$. Furthermore let $\tilde \bdelta$ represent a vector $\tilde \bdelta \equiv (\delta_0,\bdelta^\top)^\top$ for $(\delta,\bdelta) \in \bbR\times \bbR^p$. Then the RSC condition in our GLM setting becomes $\tilde \bdelta^T\calQ_\phi(t_0,\bt)\tilde \bdelta \geq \kappa \|\tilde \bdelta\|$ for all $(\delta_0,\bdelta),(t_0,\bt) \in \calC_0\times(\calC\cap \calM_0)$ for some $\kappa > 0$ (cf. Assumption 3.1 of \cite{lee2015modelselection}). In addition, the irrepresentable condition becomes $\|\calQ^{}_{\phi,M_0^c,M_0} \calQ^{-1}_{\phi,M_0,M_0}\sign(\bbeta_{M_0})\|_\infty \leq 1- \xi$ for some $\xi \in (0,1)$, where $\calQ_{,\phi,\calA,\calB}$ denotes the matrix constructed from $\calQ_\phi$ by keeping rows with indices in $\calA$ and columns with indices in $\calB$.

Now, in the idealized linear model the Hessian of the loss function $\|\bz_0 - (\bu_0 t_0 + \bU \bt)\|^2$ is exactly $\calQ_\phi$. Therefore if the RSC condition holds, then the same condition holds when $\calQ_{\phi}(t_0,\bt)$ is replaced by $\calQ_\phi = \calQ_\phi(\beta_0,\bbeta)$, since $(\beta_0,\bbeta)$ belongs to the set $\calC_0\times (\calC \cap \calM_0)$. Therefore, if the RSC is satisfied for the GLM, it will also be satisfied in the idealized linear model.  Moreover, the irrepresentable condition for the GLM is identical to its counterpart in the linear model with the idealized pseudo-data, as it is formulated in terms of Hessian evaluated at the true parameter values. 

Therefore, if these conditions are met for consistent variable selection via $L_1$-penalization of the GLM log-likelihood, they will also be met for consistent variable selection via $L_1$-penalization of the least-squares criterion in the idealized pseudo-data.  By the discussion at the end of Section \ref{sec:ideal}, model selections based on $\hat \bz_0$ and $\hat \bU_0$ will reliably match those based on the idealized $\bz_0$ and $\bU_0$, so that, by extension, model selection via $L_1$-penalization of the least-squares criterion in $\hat \bz_0$ and $\hat \bU_0$, giving $\hat M_\lambda$, will be reliable whenever model selection via $L_1$-penalization of the GLM log-likelihood, giving $\hat M_\lambda^{\text{mle}}$, is reliable. In support of these findings, we present an empirical comparison of selected models $\hat M_\lambda$ and $\hat M_\lambda^{\text{mle}}$ on simulated data sets in the Supplementary Material.

\section{Post-selection inference based on the linearized model}
\label{sec:pp}

The (centered) idealized response vector $\bz_0$ may be written as
\begin{equation}
\label{eqn:lin}
\bz_0 = \bU_0 \bbeta + \bxi_0,
\end{equation}
where the term $\bxi_0$ is defined as the centered version $\bxi_0 \equiv (\bI - \bP_0)\bxi$ of the vector $\bxi \equiv (\xi_1,\dots,\xi_n)^\top$, where $\xi_i \equiv (Y_i - b'(\eta_i))/\sqrt{b''(\eta_i)}$, $i= 1,\dots,n$ are independent random variables with mean zero and variance $a(\phi)$, so that the covariance matrix of $\bxi_0$ is $(\bI - \bP_0)a(\phi)$.

Here we apply the parametric programming (PP) approach in \citet{le2021parametric} for making post-selection inferences based on observing the idealized data $\bz_0$ and $\bU_0$ in \eqref{eqn:lin}, treating this as a Gaussian linear model.

\subsection{Post-selection sampling distribution of idealized LSE}
\label{sec:postlse}

Given a model $M\subset \{1, \ldots, p\}$ the estimator $\tilde \bbeta_M$ has entry $j$ given by $(\tilde \bbeta_M)_j =\bc_{M,j}^\top \bz_0$, and its estimation target is $(\tilde \bbeta_M^*)_j$ with $\tilde \bbeta_M^*$ defined in \eqref{eqn:betastarm}. 
Recall that $\tilde M_\lambda$ is the model selected via $L_1$-penalization in \eqref{eqn:betalambdaideal} of the least-squares criterion in the idealized pseudo-data $\bz_0$ and $\bU_0$. Given the event $\tilde M_\lambda= M$, we consider making conditional inferences on $(\tilde \bbeta_M^*)_j$ based on the conditional distribution of $(\tilde \bbeta_M)_j| \{\tilde M_\lambda= M \}$.

This is reminiscent of the problem raised in Section~\ref{sec:mleglm} of finding the distribution of $(\hat \bbeta{}^\text{mle}_M)_j|\{\hat M^\text{mle}_\lambda=M\}$ in order to make inferences on  $(\bbeta^*_M)_j$ in the GLM setting. In essence, we transform the original intractable problem in GLMs to an easier (and solved) problem in linear models. The analogy of these two problems is justified by the asymptotic equivalence of the pivotal quantities considered in Theorem \ref{thm:pivot} and, moreover, as we discuss in Section \ref{sec:glmvs}, the fact that selected models $\hat M_\lambda^\text{mle}$ and $\hat M_\lambda$ will be reliably similar under standard conditions.

Treating $\bxi_0$ as multivariate Gaussian, the estimator $(\tilde \bbeta_M)_j$, prior to conditioning on the selection of model $M$, has the $\calN( ( \tilde \bbeta_M^*)_j,a(\phi)\|\bc_{M,j}\|^2 )$ distribution, where this approximately holds when $\bxi_0$ is non-Gaussian, by Theorem 1(i). To account for variable selection, inference should be based on the \textit{conditional} distribution of $(\tilde \bbeta_M)_j$ given the selection event $\{\tilde M_\lambda=M\}=\{\bz_0\in \mathbb{R}^n:\, \tilde M_\lambda(\bz_0)=M\}$, where we use $\tilde M_\lambda(\bz_0)$ to indicate the dependence of $\tilde M_\lambda$ on $\bz_0$

Given any contrast $\bc^\top\bz_0$ of interest, the main idea of the PP approach of \cite{le2021parametric} is to ``parameterize'' all the relevant response vectors $\bz_0$ in $\mathbb{R}^n$ by linking them to a ``parameter'' $\tau$ in $\mathbb{R}$ so that the selection event can be identified with a subset of parameter values in $\mathbb{R}$ rather than with a subset of response vectors in $\mathbb{R}^n$. To achieve this parameterization, we define another event $\{\bz_0\in \mathbb{R}^n:\, \hat \bq(\bz_0)=\bq\}$, where  $\hat \bq(\bz_0) \equiv (\bI - \bP_\bc)\bz_0$, where $\bP_\bc \equiv \|\bc\|^{-2}\bc\bc^\top$, and  $\bq$ is in the column space of $(\bI - \bP_\bc)$ as a realization of $\hat \bq(\bz_0)$ in a given application where $M$ is the realization of $\tilde M_\lambda(\bz_0)$. By construction, $\hat \bq(\bz_0)$ and $\bc^\top \bz_0$ are uncorrelated, and since we are treating $\bxi_0$ as multivariate Gaussian, they are independent. Thus, using  ``$\overset{d}{=}$'' to refer to ``equivalent in distribution,'' we have
\begin{align}
&\ \bc^\top\bz_0 |\{\bz_0\in \mathbb{R}^n:\, \tilde M_\lambda(\bz_0) = M\} \nonumber\\
 \overset{d}{=} &\ \bc^\top\bz_0 | \{\bz_0\in \mathbb{R}^n:\, \tilde M_\lambda(\bz_0) = M,\, \hat \bq(\bz_0) = \bq\} \nonumber \\
\overset{d}{=} &\ \bc^\top\bz_0 | \{\bz_0\in \mathbb{R}^n:\, \tilde M_\lambda(\bz_0) = M,\, (\bI-\bP_\bc)\bz_0 = \bq\}  \nonumber\\
\overset{d}{=} &\ \bc^\top\bz_0 | \{\bz_0\in \mathbb{R}^n:\, \tilde M_\lambda(\bz_0) = M,\, \bz_0=\bq+\bP_\bc\bz_0\} \nonumber \\
\overset{d}{=} &\ \bc^\top\bz_0 | \{\bz_0\in \mathbb{R}^n:\, \tilde M_\lambda(\bq+\bP_\bc\bz_0) = M\} \nonumber \\
\overset{d}{=} &\ \bc^\top\bz_0 | \{\bz_0\in \mathbb{R}^n:\, \tilde M_\lambda(\bq+\|\bc\|^{-2}\bc\bc^\top\bz_0) = M\} \nonumber \\
\overset{d}{=} &\ \bc^\top \bz_0|\{\tau\in \mathbb{R}:\, \tilde M_\lambda(\bz_0(\tau))=M\}, \label{eqn:tauevent}
\end{align}
with $\bz_0(\tau) \equiv \bq + \tau\|\bc\|^{-2}\bc$, indexed by $\tau$, so that it moves across the support of $\bz_0$ as $\tau$ moves across $\mathbb{R}$. Similarly, we ``parameterize'' the models selected via the minimization in \eqref{eqn:betalambdaideal} be writing
\begin{equation}
\tilde \bbeta_\lambda(\tau) \equiv \argmin_{\bt \in \mathbb{R}^p} \frac{1}{2n}\|\bz_0(\tau) - \bU_0 \bt\|^2 + \lambda \|\bt\|_1,
\label{eqn:taulse}
\end{equation}
so that $\tilde M_\lambda (\bz_0(\tau)) \equiv \{j : \, (\tilde \bbeta_\lambda(\tau))_j \neq 0\}$ as $\tau$ varies in $\mathbb{R}$. This translates the search of $\bz_0$ in $\mathbb{R}^n$ for the selection event $\{ \tilde M_\lambda(\bz_0)=M\}$ to the search of $\tau$ in $\mathbb{R}$ that satisfies $\tilde M_\lambda(\bz_0(\tau))=M$ for a given model $M$.  \cite{le2021parametric} provided an efficient algorithm to identify the values of $\tau$ in  $\calT_M \equiv \{\tau \in \mathbb{R}: \, \tilde M_\lambda(\bz_0(\tau)) = M\}$ as a union of disjoint intervals in $\mathbb{R}$.  

With $\bc = \bc_{M,j}$ and $\calT_{M,j}$ as the support of the conditional distribution of $(\tilde \bbeta_M)_j = \bc_{M,j}^\top\bz_0$, by \eqref{eqn:tauevent} we have 
\[
(\tilde \bbeta{}_M)_j | \{\tilde M_\lambda(\bz_0) = M\} \sim \mathcal{N}_{\calT_{M,j}}( (\tilde \bbeta^*_M)_j, \, a(\phi) \|\bc_{M,j}\|^2),
\]
where $\mathcal{N}_{\calT_{M,j}}(a,b)$ represents the normal distribution with mean $a$ and variance $b$ truncated to have support on $\calT_{M,j}$. Denoting by $F_{\calT_{M,j}}(\cdot; a, b)$ the cumulative distribution function (CDF) of this distribution, we can pass $(\tilde \bbeta_M)_j$ through its own CDF to construct a pivotal quantity with which a post-selection $(1-\alpha)100\%$ confidence interval for $(\bbeta{}^*_M)_j$ may be constructed as
\begin{equation}
\label{eqn:ci}
\text{CI}_{M,j} \equiv \Big\{\mu\in \mathbb{R}: \, \alpha/2 \leq F_{\calT_{M,j}}( (\tilde \bbeta_M)_j; \mu,a(\phi) \|\bc_{M,j}\|^2)\leq 1 - \alpha/2\Big\}
\end{equation}
for $\alpha \in [0,1/2]$.
Similarly, a post-selection $p$-value for testing $(\tilde \bbeta^*_M)_j = 0$ versus $(\tilde \bbeta^*_M)_j \neq 0$ for $j \in M$ based on $(\tilde \bbeta_M)_j$ can be defined as
\begin{equation}
\label{eqn:pval}
2 \times \min\Big\{F_{\calT_M}((\tilde \bbeta_M)_j; 0, a(\phi)\|\bc_{M,j}\|^2), \, 1 - F_{\calT_M}((\tilde \bbeta_M)_j; 0, a(\phi)\|\bc_{M,j}\|^2)\Big\}.
\end{equation}

We propose carrying out these steps, substituting for $\bz_0$ and $\bU_0$ the observable counterparts $\hat \bz_0$ and $\hat \bU_0$ after selecting the model $\hat M_\lambda$ based on the estimator $\hat \bbeta_\lambda$ in \eqref{eqn:betalambda}.

\subsection{Accounting for penalty parameter selection}
\label{sec:postlamb}
Instead of being pre-specified as assumed in Section~\ref{sec:postlse}, we may allow the penalty parameter $\lambda$ to be selected based on the data $\bz_0$ and $\bU_0$, as is usually done in practice. This change demands a revision of the selection event in \eqref{eqn:tauevent} to acknowledge the additional selection of $\lambda$. This extra selection can again be ``parameterized'' via $\tau$ when $\lambda$ is chosen based on one partition of $(\bz_0, \bU_0)$ into a training set, $(\bz_0^{\text{train}}, \bU_0^{\text{train}})$, and a validation set, $(\bz_0^{\text{val}}, \bU_0^{\text{val}})$. 

More specifically, the selection of $\lambda$ can be formulated as an  optimization indexed by $\tau$, $
\tilde \lambda(\bz_0(\tau)) \equiv \argmin_{\lambda \in \Lambda} \|\bz_0^{\text{val}}(\tau) - \bU_0^{\text{val}}\tilde{\bbeta}{}_\lambda^{\text{train}}(\bz(\tau))\|^2$, 
where $\Lambda$ is the set of candidate values for $\lambda$, and, for each $\lambda\in \Lambda$, mimicking \eqref{eqn:taulse}, $
\tilde{\bbeta}{}_\lambda^{\text{train}}(\bz_0(\tau)) \equiv \argmin_{\bt\in\mathbb{R}^p}(2n)^{-1}\|\bz_0^{\text{train}}(\tau) - \bU^{\text{train}}_0\bt\|^2 + \lambda\|\bt\|_1$. Define $\calT_\lambda \equiv \{\tau\in \mathbb{R}:\, \tilde \lambda(\bz_0(\tau))=\lambda\}$, viewing $\lambda$ as a realization of $\tilde \lambda(\bz_0(\tau))$ corresponding to $M$ as the realization of $\tilde M_\lambda(\bz_0(\tau))$. Then the complete selection event is  $\calT_\lambda\cap \calT_{M,j} \equiv \{\tau\in \mathbb{R}:\, \tilde \lambda(\bz_0(\tau))=\lambda, \, \tilde M_\lambda(\bz_0(\tau))=M\}$, and thus $(\tilde \bbeta_M)_j|\{\tilde M_\lambda=M\}\sim \mathcal{N}_{\calT_\lambda\cap \calT_{M,j}}((\tilde \bbeta^*_M)_j, \, a(\phi)\|\bc_{M,j}\|^2)$. \cite{le2021parametric} developed an algorithm for identifying $\calT_\lambda$. It is now  straightforward to obtain a $(1-\alpha)100\%$ confidence interval for $(\tilde \bbeta^*_M)_j$ and to compute the $p$-value for testing the significance of this covariate effect based on $(\tilde \bbeta_M)_j$: one simply changes the support of the post-selection sampling distribution of $(\tilde \bbeta_M)_j$ from $\calT_{M,j}$ to $\calT_\lambda \cap \calT_{M,j}$ in the distribution function in \eqref{eqn:ci} and \eqref{eqn:pval}.

\section{Implementation of the proposed method}
\label{sec:3models}

Here we describe the construction of $\hat \bz_0$ and $\hat \bU_0$ in three non-Gaussian models. The first two are GLMs, while the third, though not a GLM, admits a construction of $\hat \bz_0$ and $\hat \bU_0$ analogous to that in GLMs. We note that if the MLE is undefined due to, for example, a complete or quasi-complete separation in binary response data \citep{albert1984existence}, or if $p > n$, we prescribe replacing the MLE with a slightly regularized estimator, such as an $L_1$-penalized estimator with weak penalization. Theorem 1(ii) will hold as long as $\hat \bz_0$ and $\hat \bU_0$ are constructed with a consistent estimator of $(\beta_0,\bbeta)$.

\subsection{Logistic and Poisson models}
\label{sec:2glms}
For a binary response, the logistic regression model is most widely used. Here $a(\phi) = 1$ and $b(\eta)= \log(1 + e^{\eta})$ for all $\eta \in \bbR$ in \eqref{eqn:glm}, yielding $b'(\eta) = e^{\eta}/(1 + e^{\eta})$ and $b''(\eta) = b'(\eta)(1 - b'(\eta))$ as the key quantities needed to evaluate the $\hat \bz_0$ and $\hat \bU_0$. For responses which are counts, the Poisson regression model is often used. Here $a(\phi)=1$ and $b(\eta) = b'(\eta) = b''(\eta) = e^{\eta}$ for all $\eta \in \bbR$.  If a GLM involves an unknown $\phi$, we prescribe substituting for $\phi$ the MLE $\hat \phi$. This is also our strategy for dealing with a nuisance parameter irrelevant to the linear predictor $\eta$ in the third regression model described next.

\subsection{Beta regression}
\label{sec:betareg}

For continuous responses $Y_1,\dots,Y_n \in [0, 1]$, such as rates or proportions, observed with $\bx_1,\dots,\bx_n \in \bbR^p$, \cite{ferrari2004beta} considered beta regression under which $Y_i \sim \text{Beta}(\mu_i\phi,\, \nu_i\phi)$
with $\mu_i \equiv 1/(1+e^{-\eta_i})$ and $\nu_i \equiv 1-\mu_i$ and $\eta_i$ as before, for $i=1,\dots,n$, where $\phi > 0$ is a precision parameter.  In this model $Y_i$ has mean $\mu_i$ and variance $\mu_i(1-\mu_i)/(1+\phi)$ for $i = 1,\dots,n$. The NR update is not the same in this setting as in Section \ref{sec:glmlinear} due to the fact that the Hessian of the log-likelihood depends on the responses \citep[see Appendix A in][]{ferrari2004beta}.
As an alternative to NR, we adopt Fisher scoring \citep[][Section 4.4.1]{davison2003statistical}, whereby the Hessian is replaced by its expected value. If $\phi$ is known, we can describe the Fisher scoring update by setting  $\mu_i(t_0,\bt) \equiv 1/(1 + e^{-\eta_i(t_0,\bt)})$ and $\nu_i(t_0,\bt) \equiv 1 - \mu_i(t_0,\bt)$ with $\eta_i(t_0,\bt) \equiv t_0 + \bx_i^\top \bt$ as well as $\tilde Y_i \equiv \log(Y_i/(1-Y_i))$ and $\tilde \mu_i(t_0,\bt) \equiv \psi(\mu_i(t_0,\bt) \phi)-\psi(\nu_i(t_0,\bt)\phi)$ for $i=1,\dots,n$, where $\psi(\cdot)$ is the digamma function. From here we define weights
\[
w_i(t_0,\bt) \equiv \sqrt{\phi}  \mu_i(t_0,\bt) \nu_i(t_0,\bt) \sqrt{\psi'( \mu_i(t_0,\bt) \phi) + \psi'(\nu_i(t_0,\bt) \phi)}
\]
for $i=1,\dots,n$ and the vector $\bz(t_0,\bt)$ having entries
\[
z_i(t_0,\bt) = w_i(t_0,\bt) \eta_i(t_0,\bt) + \mu_i(t_0,\bt)\nu_i(t_0,\bt)(\tilde Y_i - \tilde \mu(t_0,\bt))/w_i(t_0,\bt),
\]
the vector $\bu_0(t_0,\bt)$ having entries $w_i(t_0,\bt)$ and the matrix $\bU(t_0,\bt)$ having rows $w_i(t_0,\bt)\bx_i^\top$ for $i=1,\dots,n$. Then, for a fixed value of $\phi$, an initial value $(t_0^{(0)},\bt^{(0)})$ can be updated as in \eqref{eqn:nrls}.

To estimate unknown $\phi$, one maximizes the log-likelihood evaluated at the current $(t_0^{(k)}, \bt^{(k)})$ with respect to $\phi$ (the maximizer can be obtained in closed form); then one updates $(t_0^{(k)}, \bt^{(k)})$ via \eqref{eqn:nrls} with $\phi$ fixed at its current estimate. Iterating until convergence yields the MLE $(\hat \beta_0,\hat \bbeta,\hat \phi)$ of $(\beta_0,\bbeta,\phi)$. As before, we set $\hat \bz \equiv \bz(\hat \beta_0,\hat \bbeta)$, $ \hat \bu_0 \equiv \bu_0(\hat \beta_0,\hat\bbeta)$, and $\hat \bU \equiv \bU(\hat \beta_0,\hat \bbeta)$ as the pseudo-data, and to disregard the intercept we construct $\hat \bz_0$ and $\hat \bU_0$ as in Section \ref{sec:glmlinear}.

The idealized counterparts to $\hat \bz$, $ \hat \bu_0$, and $\hat \bU$ are defined, as before, as $\bz\equiv \bz(\beta_0,\bbeta)$, $\bu_0 \equiv \bu_0(\beta_0,\bbeta)$, and $\bU \equiv \bU(\beta_0,\bbeta)$, and the ``centered versions'' are also obtained as before. Thus we may write $\bz_0 = \bU_0\bbeta + \bxi_0$ as in \eqref{eqn:lin}, where $\bxi_0 = (\bI - \bP_0)\bxi$, except that in beta regression the vector $\bxi$ has entries $\xi_i \equiv w_i^{-1} \mu_i \nu_i(\tilde Y_i - \tilde \mu_i)$ with $w_i \equiv w_i(\beta_0,\bbeta)$ and $\tilde \mu_i \equiv \tilde \mu_i(\beta_0,\bbeta)$ for $i=1,\dots,n$. Since $\tilde Y_i$ has mean $\tilde \mu_i$ and variance $\psi'(\mu_i \phi)+\psi'(\nu_i\phi)$, $\xi_i$ has mean zero and variance $\phi^{-1}$ for $i=1,\dots,n$. The covariance matrix of $\bxi_0$ is thus $(\bI - \bP_0)\phi^{-1}$ in beta regression. Therefore, in Section \ref{sec:postlse} $a(\phi)$ is replaced with $a(\phi) \equiv \phi^{-1}$, where for the value of $\phi$ we plug in the MLE.

\section{Simulation study}
\label{sec:sim}

Here compare on simulated data sets the performance of the proposed method of applying parametric programming after linearization, which we abbreviate as PPL, with i) standard Wald-type inference in the selected model without any conditioning on the selection event and with ii) polyhedral-based post-selection inference as in \cite{lee2016exact} following out linearization step. Comparisons of PPL and data-splitting are provided in the Supplementary Material.

\subsection{Parametric programming versus the naive method}
\label{sec:simu1}
Under the logistic, Poisson, and beta regression models in Section~\ref{sec:3models}, we generate response data after drawing covariate vectors $\bx_1,\dots,\bx_n \in \bbR^p$ having independent $\mathcal{N}(0, 1)$ entries under $n=500$ and $p=20$. On each of 1000 simulated data sets we implement our PPL method as well as the naive method to construct confidence intervals for the coefficients in the selected model $\hat M_\lambda$. On each data set we perform model selection at each value in a grid of twenty values for $\lambda$, equally spaced on logarithmic scale. In each model we set $\beta_0 = -2$ and $M_0 = \{1,2,3\}$. For logistic regression we set $\bbeta_{M_0} = (2,2,1)^\top$, for Poisson, $\bbeta_{M_0} = (1,1,-1)^\top$, and for beta, $\bbeta_{M_0} = (1,-1/2,1/2)^\top$ and consider a set of $\lambda$ values in the intervals $[2,12]$, $[8,56]$, and $[2,10]$, respectively, where these intervals were chosen to yield wide spreads of model sizes.

For each simulated data set, at each $\lambda$, we perform post-selection inference by constructing the confidence interval $\text{CI}_{M,j}$ in \eqref{eqn:ci} at $\alpha = 0.05$ for each index $j$ in the selected model $M$. We then record a realization of the Type I error rate as the proportion of unimportant covariates in $M$ for which a nonzero regression coefficient is inferred, that is, we record the ratio  
$$
\text{Type I error} \equiv \frac{|\{j\in M:  \, 0 \notin \text{CI}_{M,j} \text{ and } (\bbeta_M)_j=0 \}|}{|\{j\in M: \, (\bbeta_M)_j=0\}|},
$$
where $\text{CI}_{M,j}$ is defined in \eqref{eqn:ci}.  If $|\{j\in M: \, \beta_j=0\}|=0$, we record a zero for the Type I error. For the naive method, we use Wald-type confidence intervals based on $\hat \bbeta{}^{\text{mle}}_M$ that ignore selection events.

\begin{figure}
    \vspace{-0.3cm}
    \centering
    \begin{tabular}{c}
        \vspace{-0.3cm}
        \includegraphics[width=70mm]{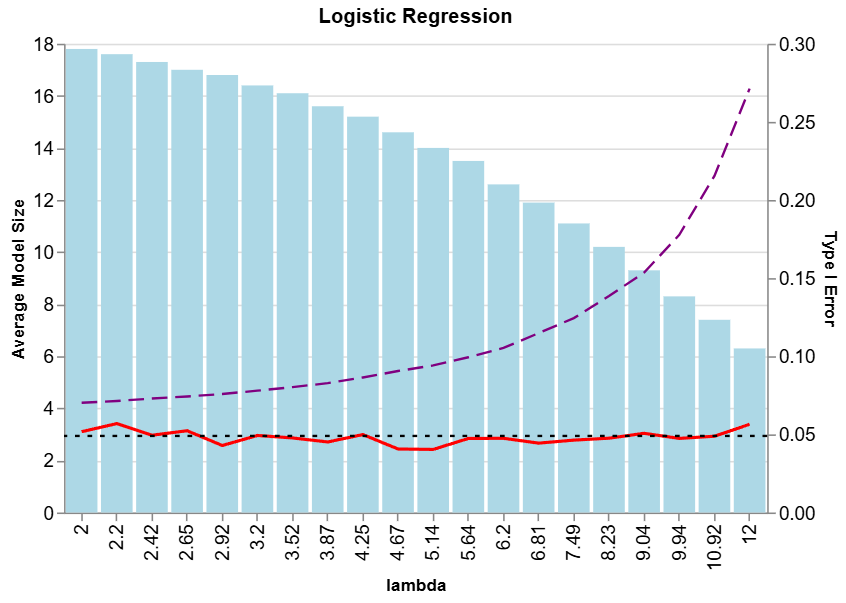} \\
        \vspace{-0.3cm}
        \includegraphics[width=70mm]{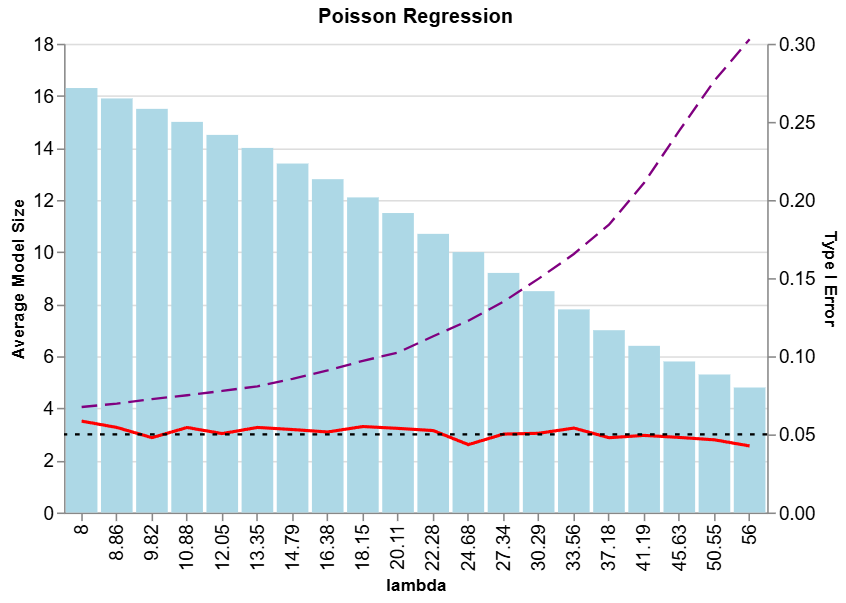} \\
        \vspace{-0.5cm}
        \includegraphics[width=70mm]{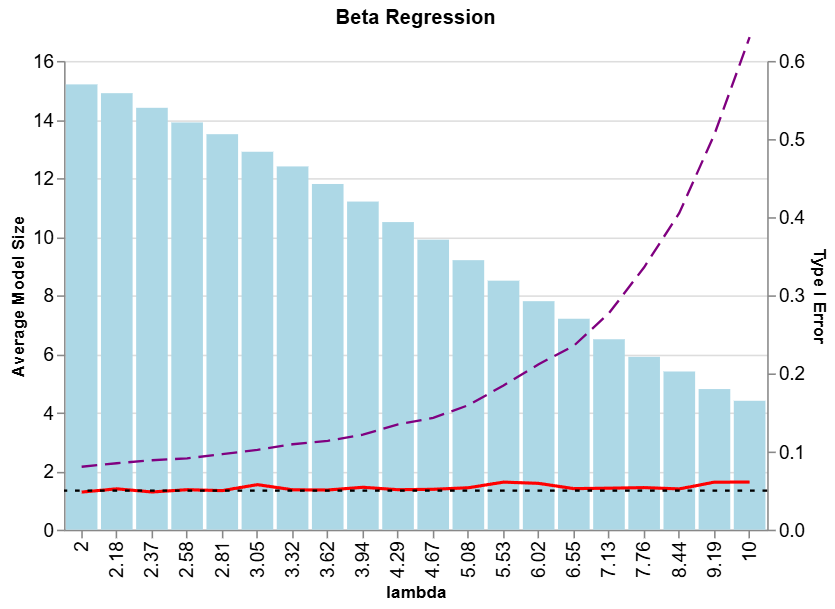}
    \end{tabular}
    \caption{Average Type I error across $1000$ Monte Carlo replicates over the grid of $\lambda$ values for logistic, Poisson, and beta regression achieved by the PPL method (solid lines) and by the naive method (dashed lines). Heights of bars indicate average sizes of the selected models across the $\lambda$ values. }
    \label{fig:typeIplots}
\end{figure}

Figure~\ref{fig:typeIplots} shows the average Type I error achieved by the PPL method across 1000 Monte Carlo replicates for all regression settings at each $\lambda$ with the average Type I error from the naive method overlaid. The average sizes of the selected models at each $\lambda$ are also depicted (by the heights of the bars). As expected, the naive method leads to increasingly inflated Type I error rates as $\lambda$ increases. In contrast, our PPL method provides reliable inference after model selection, maintaining a Type I error close to the nominal level of $0.05$ across all $\lambda$ values.

We also compare the average Type I error achieved by the PPL and naive methods under data-based selection of the tuning parameter $\lambda$. Here we select a value of $\lambda$ from among the grid of twenty values as described in Section \ref{sec:postlamb} using a single $70\%$/$30\%$ split of the data. For logistic regression, the average Type I error rates achieved by the PPL and naive methods under $\alpha = 0.05$ were $0.049$ and $0.198$, respectively; for Poisson regression these were $0.044$ and $0.269$, respectively; and for beta regression these were $0.048$ and $0.349$ respectively.  Here also we see that the proposed method maintains an average Type I error close to $0.05$ across all three regression settings, whereas the naive method leads to drastically inflated Type I errors. This also showcases the versatility of the PPL method when compared with the polyhedral method, which cannot accommodate data-driven penalty parameter selection. Even without involving the additional selection of $\lambda$, our method still outperforms the polyhedral method, as we show next.



\subsection{Parametric programming versus polyhedral method}
\label{sec:simu2}
Our method of parametric programming after linearization (PPL) prescribes applying the PP approach to a linear model following a linearization step.  Alternatively, following the linearization step, one could apply the polyhedral method described in \cite{lee2016exact}.  The advantages of the PP method over the polyhedral method are twofold. First, the PP method avoids the ``overconditioning'' (on the signs of selected coefficients) that the polyhedral method involves.  The minimal conditioning of the PP method enhances the statistical efficiency compared to that of the polyhedral method. Second, the PP method can easily account for data-driven selection of $\lambda$.

Focusing on logistic regression, we compare in Table~\ref{table:PPvsAdj} the performances of $95\%$ confidence intervals based on these two strategies. As two valid post-selection inference methods, they both preserve coverage probabilities close to the nominal level of 95\%. For the truly non-zero coefficient $\beta_2$, the average widths of the CIs obtained from the PP method tend to be wider than those obtained from the polyhedral method for most $\lambda$ values. But, for $\beta_4$ and $\beta_6$, which are truly equal to zero, the CIs from the PP method are substantially tighter than those from the polyhedral method. This indicates that the over-conditioning (on signs) required by the polyhedral method is particularly detrimental to statistical power when inferring a null covariate effect, as the signs of these regression coefficients in the estimated model can be positive or negative with non-vanishing probability even as the sample size grows. In contrast, coefficients that are far away from zero will, with high probability, have the correct signs in the selected model, so conditioning on their signs has less impact on statistical power.

\begingroup

\renewcommand{\arraystretch}{0.6} 
\begin{table}[h]
\centering
\caption{\label{table:PPvsAdj}Average lower and upper bounds, average width, and empirical coverage probability of 95\% confidence intervals from the PPL method and the polyhedral method at certain $\lambda$ values.}
\resizebox{\textwidth}{!}{
\setlength\tabcolsep{3pt}
\begin{tabular}{cccccccc}
\hline
\multirow{2}{*}{Coefficient} & \multicolumn{3}{c}{PPL} &&   \multicolumn{3}{c}{Polyhedral} \\ 
& $95\%$ CI & Width & Coverage && $95\%$ CI & Width & Coverage \\ 
\cline{1-4} \cline{6-8} 
& \multicolumn{7}{c}{$\lambda=2$}\\
$\beta_2$ & {[}$0.50$, $1.52${]} & $1.02$ & $93.0$ && {[}$0.54$, $1.61${]} & $1.07$ & $93.5$ \\
$\beta_4$ & {[}$-0.49$, $0.54${]} & $1.03$ & $95.8$ && {[}$-0.74$, $1.33${]} & $2.07$ & $94.8$ \\
$\beta_6$ & {[}$-0.50$, $0.52${]} & $1.02$ & $94.9$ && {[}$-0.74$, $1.35${]} & $2.07$ & $95.1$ \\ 
\cline{1-4} \cline{6-8}
& \multicolumn{7}{c}{$\lambda=3.5$}\\
$\beta_2$ & {[}$1.36$, $2.71${]} & $1.35$ & $95.8$ && {[}$1.57$, $2.70${]} & $1.13$ & $90.5$ \\
$\beta_4$ & {[}$-0.42$, $0.44${]}& $0.86$ & $94.8$ && {[}$-0.62$, $0.61${]} & $1.23$& $93.8$ \\
$\beta_6$ & {[}$-0.44$, $0.43${]}& $0.87$ & $96.6$ && {[}$-0.61$, $0.57${]} & $1.18$& $94.9$ \\ 
\cline{1-4} \cline{6-8}
& \multicolumn{7}{c}{$\lambda=8.2$}\\
$\beta_2$ & {[}$1.13$, $2.57${]} & $1.44$ & $96.6$ && {[}$1.56$, $2.53${]} & $0.97$& $93.6$ \\
$\beta_4$ & {[}$-0.41$, $0.46${]} & $0.87$ & $94.9$ && {[}$-0.66$, $0.59${]} & $1.25$& $94.7$ \\
$\beta_6$ & {[}$-0.49$, $0.42${]}& $0.91$  & $95.5$ && {[}$-0.61$, $0.59${]} & $1.20$ & $97.6$ \\ 
\cline{1-4} \cline{6-8}
& \multicolumn{7}{c}{$\lambda=12$}\\
$\beta_2$ & {[}$1.20$, $2.47${]} & $1.27$ & $96.5$ && {[}$1.53$, $2.42${]} & $0.89$ & $93.5$ \\
$\beta_4$ & {[}$-0.42$, $0.47${]} & $0.89$ & $94.6$ && {[}$-0.56$, $0.57${]} & $1.13$ & $93.8$ \\
$\beta_6$ & {[}$-0.44$, $0.43${]} & $0.87$ & $93.5$ && {[}$-0.58$, $0.62${]} & $1.20$ & $96.9$ \\ \hline
\end{tabular}}
\end{table}
\endgroup

\section{Illustrations on real data sets}
\label{sec:real}

\subsection{Logistic regression for binary data}

The Spambase data set from the UCI Machine Learning Repository contains descriptive information for $n=4601$ emails, each classified as spam or not spam \citep{spambase_94}.  There are $p=57$ numeric features giving, for example, the frequencies of certain words or symbols, or the lengths of sequences of capital letters.  We fit a logistic regression model to predict whether to classify an email as spam.  After selecting $\lambda$ using 70\% of the data as training data and 30\% as validation data, $25$ variables were selected.  Our PPL method for post-selection inference found $13$ of the variables to be significant in classifying emails as spam or not spam, while naive inference found 19 of the variables to be significant at the 0.05  significance level. 

Table \ref{tab:logistic_real} summarizes the results, showing confidence intervals from the PPL method and the naive method, along with $p$-values associated with the $19$ variables deemed significant according to the latter. The proposed method produces CIs generally wider than those from the naive method, and with larger $p$-values than their naive counterparts. All selected features fall in the three types of features frequently reported as strong predictors in existing studies \citep{Prerika}: word frequency features, capital letter features, and special character frequency features. Among the six selected features identified as statistically significant by the naive method but deemed insignificant by the PPL method, three ({\it word\_freq\_all}, {\it word\_freq\_will}, {\it word\_freq\_email}) are not among the top features selected by most existing studies.
\begingroup

\renewcommand{\arraystretch}{0.6} 
\begin{table}[h]
    \centering
\caption{  \label{tab:logistic_real} Inferences on selected covariate effects for the Spambase data set for the 19 covariates found to be significant according to the naive method. Names of covariates found insignificant by PPL are italicized.}
    \resizebox{\textwidth}{!}{
    \begin{tabular}{l |c c c |c c c}
    \hline
        & \multicolumn{3}{c|}{\textbf{PPL}} & \multicolumn{3}{c}{\textbf{Naive}}\\
        \textbf{Covariate} & $p$-value & CI & Width & $p$-value & CI & Width \\
        \hline
        \textit{word\_freq\_all} & $0.32$ & {[}$-0.11$, $0.20${]} & $0.31$ & $0.02$ & {[}$0.02$, $0.20${]} & $0.18$ \\
        word\_freq\_our  & $0.00$ & {[}$0.22$, $0.49${]} & $0.26$ & $0.00$ & {[}$0.23$, $0.40${]} & $0.17$ \\
        word\_freq\_over & $0.02$ & {[}$0.04$, $0.32${]} & $0.28$ & $0.00$ & {[}$0.11$, $0.31${]} & $0.19$ \\
        word\_freq\_remove  & $0.00$ & {[}$0.54$, $0.96${]} & $0.42$ & $0.00$ & {[}$0.84$, $1.25${]} & $0.41$ \\
        word\_freq\_internet  & $0.00$ & {[}$0.11$, $0.38${]} & $0.27$ & $0.00$ & {[}$0.21$, $0.43${]} & $0.23$ \\
        \textit{word\_freq\_will} & $0.18$ & {[}$-0.34$, $0.09${]} & $0.43$ & $0.00$ & {[}$-0.39$, $-0.15${]} & $0.24$ \\
        word\_freq\_free  & $0.00$ & {[}$0.53$, $0.95${]} & $0.42$ & $0.00$ & {[}$0.40$, $0.69${]} & $0.30$ \\
        word\_freq\_business  & $0.01$ & {[}$0.08$, $0.43${]} & $0.34$ & $0.00$ & {[}$0.18$, $0.41${]} & $0.24$ \\
        \textit{word\_freq\_email} & $0.21$ & {[}$-0.06$, $0.26${]} & $0.33$ & $0.01$ & {[}$0.03$, $0.23${]} & $0.20$ \\
        word\_freq\_you  & $0.00$ & {[}$0.05$, $0.32${]} & $0.27$ & $0.00$ & {[}$0.31$, $0.52${]} & $0.21$ \\
        \textit{word\_freq\_your} & $0.14$ & {[}$-0.12$, $0.42${]} & $0.54$ & $0.00$ & {[}$0.27$, $0.47${]} & $0.20$ \\
        word\_freq\_000  & $0.00$ & {[}$0.36$, $0.82${]} & $0.47$ & $0.00$ & {[}$0.60$, $1.07${]} & $0.47$ \\
        word\_freq\_money  & $0.00$ & {[}$0.10$, $0.38${]} & $0.28$ & $0.00$ & {[}$0.24$, $0.63${]} & $0.39$ \\
        word\_freq\_re  & $0.05$ & {[}$-1.13$, $-0.00${]} & $1.02$ & $0.00$ & {[}$-1.18$, $-0.62${]} & $0.55$ \\
        \textit{word\_freq\_edu} & $0.80$ & {[}$-1.59$, $3.68${]} & $5.27$ & $0.00$ & {[}$-1.66$, $-0.72${]} & $0.95$ \\
        char\_freq\_!  & $0.00$ & {[}$0.24$, $0.53${]} & $0.29$ & $0.00$ & {[}$0.54$, $0.90${]} & $0.36$ \\
        char\_freq\_\$  & $0.00$ & {[}$0.56$, $0.98${]} & $0.42$ & $0.00$ & {[}$0.70$, $1.03${]} & $0.33$ \\
        capital\_run\_length\_longest  & $0.05$ & {[}$0.00$, $0.30${]} & $0.29$ & $0.00$ & {[}$0.35$, $0.66${]} & $0.31$ \\
        \textit{capital\_run\_length\_total} & $0.32$ & {[}$-0.20$, $0.36${]} & $0.56$ & $0.00$ & {[}$0.08$, $0.29${]} & $0.21$ \\
        \hline
    \end{tabular}}    
\end{table}
\endgroup

\subsection{Poisson regression for count data}
Here, we analyze the Medicaid1986 data from the R package \texttt{AER} \citep{AERpackage} using Poisson regression. These are Medicaid utilization data from the 1986 Medicaid Consumer Survey. Considering only those records under the Aid to Families with Dependent Children program and encoding categorical variables with indicators, the data set contains $n=485$ observations and $p=12$ covariates. The response is the number of doctor visits.

After selecting the penalty parameter $\lambda$ with $70\%$ of the data used for training data and $30\%$ used as validation data, we select $4$ covariates. Naive post-selection inference finds all four significantly associated with the number of doctor visits; however, our post-selection inference procedure finds only three to be significant.  Table \ref{tab:poi_real} summarizes the results, showing that the naive method claims stronger covariate effects after model selection than our PPL method. For example,  the $p$-value associated with the covariate {\it school} from the PPL method is above 0.05, while the naive method produces a nearly zero $p$-value. Previous analyses on the data \citep{gurmu1997semi} confirmed that health status measures, such as {\it health1}, are among the most important predictors for the number of doctor visits; covariates such as {\it school} become less important once these measures are included in the model.

\begingroup

\renewcommand{\arraystretch}{0.6} 
\begin{table}[h]
    \centering
    \caption{Inferences on selected covariate effects for the Medicaid data set. Names of covariates found insignificant by PPL are italicized.}
    \label{tab:poi_real}
    \resizebox{\textwidth}{!}{
\begin{tabular}{l |c c c |c c c}
\hline
        & \multicolumn{3}{c|}{\textbf{PPL}} & \multicolumn{3}{c}{\textbf{Naive}}\\
        \textbf{Covariate} & $p$-value & CI & Width & $p$-value & CI & Width \\
        \hline
        children & $0.01$ & {[}$-0.27$, $-0.06${]} & $0.21$ & $0.00$ & {[}$-0.27$, $-0.11${]} & $0.16$ \\
        health1  & $0.00$ & {[}$0.36$, $0.50${]} & $0.14$ & $0.00$ & {[}$0.36$, $0.47${]} & $0.11$ \\
        access   & $0.01$ & {[}$0.06$, $1.35${]} & $1.29$ & $0.00$ & {[}$0.08$, $0.22${]} & $0.14$ \\
        \textit{school}   & $0.07$ & {[}$-0.02$, $0.26${]} & $0.28$ & $0.00$ & {[}$0.10$, $0.26${]} & $0.16$ \\
        \hline
    \end{tabular}}    
\end{table}
\endgroup

\subsection{Beta regression for proportion data}
Lastly, we perform beta regression on the Student Performance data set from the UCI Machine Learning Repository \citep{studentperformancedata}, which examines student achievement at two Portuguese secondary schools. Covariate information includes students' grades and demographic, social, and school-related attributes. After encoding categorical variables and removing outliers, the data set contains $n=649$ observations and $p=39$ covariates. The response variable is a student's final grade (ranging from 0 to 20) divided by 20.

Using beta regression, after selecting $\lambda$ with $70\%$ of the data used as training data and $30\%$ as validation data, $27$ covariates were included in the selected model. Our post-selection method identifies $10$ variables as significant, while the naive method identifies $13$ variables. Table \ref{tab:beta_real} displays the inference results from the two methods. After adjusting for model selection, the PPL method is less aggressive in claiming the strength of covariate effects, with wider interval estimates and larger $p$-values than those from the naive method. In particular, the PPL method concludes that the covariates, {\it goout} and {\it health}, are statistically insignificant, although they are significant at the 0.05 significance level according to the naive method. The findings of the PPL method are more in line with the consensus in the literature \citep{bhatia2025comparative, kesgin2025beyond}: among the non-grade features, the covariate {\it absences} and {\it failures} are top-ranked features, study time and academic support matter (e.g., {\it studytime}, {\it schoolsup\_yes}, {\it higher\_yes}) also tend to be highly influential on a student's final grade, parental education ({\it Medu}) is moderately influential, whereas some social and lifestyle variables, such as {\it goout} and {\it health}, have much weaker predictive power.

\begingroup

\renewcommand{\arraystretch}{0.6} 
\begin{table}[H]
    \centering
    \caption{\label{tab:beta_real} Inferences on selected covariate effects for the student performance data set for the 13 covariates found to be significant according to the naive method. Names of covariates found insignificant by PPL are italicized.}
    \resizebox{\textwidth}{!}{
    \begin{tabular}{l |c c c |c c c}
    \hline
        & \multicolumn{3}{c|}{\textbf{PPL}} & \multicolumn{3}{c}{\textbf{Naive}}\\
        \textbf{Covariate} & $p$-value & CI & Width & $p$-value & CI & Width \\
        \hline
        age  & $0.00$ & {[}$0.04$, $0.13${]} & $0.09$ & $0.00$ & {[}$0.04$, $0.13${]} & $0.08$ \\
        Medu  & $0.02$ & {[}$0.02$, $0.23${]} & $0.21$ & $0.01$ & {[}$0.02$, $0.11${]} & $0.10$ \\
        studytime  & $0.02$ & {[}$0.01$, $0.10${]} & $0.09$ & $0.01$ & {[}$0.02$, $0.10${]} & $0.08$ \\
        failures  & $0.00$ & {[}$-0.19$, $-0.10${]} & $0.09$ & $0.00$ & {[}$-0.19$, $-0.10${]} & $0.08$ \\
        \textit{goout} & $0.17$ & {[}$-0.11$, $0.02${]} & $0.14$ & $0.04$ & {[}$-0.09$, $-0.00${]} & $0.09$ \\
        \textit{health} & $0.14$ & {[}$-0.08$, $0.02${]} & $0.10$ & $0.03$ & {[}$-0.08$, $-0.00${]} & $0.08$ \\
        absences  & $0.00$ & {[}$-0.14$, $-0.03${]} & $0.10$ & $0.00$ & {[}$-0.12$, $-0.04${]} & $0.08$ \\
        school\_MS  & $0.03$ & {[}$-0.12$, $-0.00${]} & $0.11$ & $0.00$ & {[}$-0.11$, $-0.02${]} & $0.08$ \\
        sex\_M  & $0.02$ & {[}$-0.11$, $-0.01${]} & $0.10$ & $0.00$ & {[}$-0.11$, $-0.02${]} & $0.09$ \\
        Fjob\_teacher  & $0.03$ & {[}$0.01$, $0.10${]} & $0.09$ & $0.01$ & {[}$0.02$, $0.10${]} & $0.08$ \\
        \textit{reason\_reputation} & $0.13$ & {[}$-0.02$, $0.10${]} & $0.12$ & $0.02$ & {[}$0.01$, $0.09${]} & $0.08$ \\
        schoolsup\_yes  & $0.00$ & {[}$-0.12$, $-0.04${]} & $0.08$ & $0.00$ & {[}$-0.12$, $-0.04${]} & $0.08$ \\
        higher\_yes  & $0.00$ & {[}$0.07$, $0.14${]} & $0.08$ & $0.00$ & {[}$0.07$, $0.15${]} & $0.08$ \\
        \hline
    \end{tabular}}    
\end{table}
\endgroup

\section{Discussion}
\label{sec:disc}

We propose parametric programming following a linearization step for performing post-selection inference in generalized linear models, which can be adapted to models outside of GLMs, such as the beta regression model. The proposed method addresses key limitations of the polyhedral method, which involves over-conditioning, leading to wider confidence intervals and compromised statistical power. The proposed method can also adjust for data-based penalty parameter selection, in contrast to the polyhedral method, which assumes a fixed penalty parameter chosen prior to the observation of the data.

Compared to the work by \citet{taylor2018post}, where extension of the polyhedral method in generalized regression models was considered, our work provides more insight into inferences and variable selection based on the pseudo-data in a linear model and those based on the original data in a GLM. The gained insight can potentially lead to further extension of the proposed methodology in several interesting directions. These include post-selection inference in nonparametric regression models, or when response data are partially observed and prone to error as in group testing settings, or when covariates are prone to measurement error. Our current approach is grounded in maximum likelihood estimation. Generalizing it to the broader M-estimation framework is another reachable goal that can broaden its applicability. 

Computer code for implementing our proposed method and competing methods considered in the simulation study are available at \url{https://github.com/kateshen28/InfGLM}.


\par


\bibhang=1.7pc
\bibsep=2pt
\fontsize{9}{14pt plus.8pt minus .6pt}\selectfont
\renewcommand\bibname{\large \bf References}
\expandafter\ifx\csname
natexlab\endcsname\relax\def\natexlab#1{#1}\fi
\expandafter\ifx\csname url\endcsname\relax
  \def\url#1{\texttt{#1}}\fi
\expandafter\ifx\csname urlprefix\endcsname\relax\def\urlprefix{URL}\fi

\bibliographystyle{chicago}      
\bibliography{refs}   

\begin{thebibliography}{}

\bibitem[\protect\citeauthoryear{Albert and Anderson}{Albert and Anderson}{1984}]{albert1984existence}
Albert, A. and J.~A. Anderson (1984).
\newblock On the existence of maximum likelihood estimates in logistic regression models.
\newblock {\em Biometrika\/}~{\em 71\/}(1), 1--10.

\bibitem[\protect\citeauthoryear{Bachoc, Leeb, and P{\"o}tscher}{Bachoc et~al.}{2019}]{bachoc2019valid}
Bachoc, F., H.~Leeb, and B.~M. P{\"o}tscher (2019).
\newblock Valid confidence intervals for post-model-selection predictors.
\newblock {\em The Annals of Statistics\/}~{\em 47\/}(3), 1475--1504.

\bibitem[\protect\citeauthoryear{Bachoc, Preinerstorfer, and Steinberger}{Bachoc et~al.}{2020}]{bachoc2020uniformly}
Bachoc, F., D.~Preinerstorfer, and L.~Steinberger (2020).
\newblock Uniformly valid confidence intervals post-model-selection.
\newblock {\em The Annals of Statistics\/}~{\em 48\/}(1), 440--463.

\bibitem[\protect\citeauthoryear{Berk, Brown, Buja, Zhang, Zhao, et~al.}{Berk et~al.}{2013}]{berk2013valid}
Berk, R., L.~Brown, A.~Buja, K.~Zhang, L.~Zhao, et~al. (2013).
\newblock Valid post-selection inference.
\newblock {\em The Annals of Statistics\/}~{\em 41\/}(2), 802--837.

\bibitem[\protect\citeauthoryear{Bhatia, Yadav, Sharma, Shubneet, Yadav, and Talwandi}{Bhatia et~al.}{2025}]{bhatia2025comparative}
Bhatia, R., S.~Yadav, R.~Sharma, Shubneet, A.~R. Yadav, and N.~S. Talwandi (2025).
\newblock A comparative study of feature selection techniques for predicting student academic performance using educational data.
\newblock In {\em International Conference on Data Analytics \& Management}, pp.\  352--362. Springer.

\bibitem[\protect\citeauthoryear{Davison}{Davison}{2003}]{davison2003statistical}
Davison, A.~C. (2003).
\newblock {\em Statistical models}, Volume~11.
\newblock Cambridge university press.

\bibitem[\protect\citeauthoryear{Ferrari and Cribari-Neto}{Ferrari and Cribari-Neto}{2004}]{ferrari2004beta}
Ferrari, S. and F.~Cribari-Neto (2004).
\newblock Beta regression for modelling rates and proportions.
\newblock {\em Journal of applied statistics\/}~{\em 31\/}(7), 799--815.

\bibitem[\protect\citeauthoryear{Gurmu}{Gurmu}{1997}]{gurmu1997semi}
Gurmu, S. (1997).
\newblock Semi-parametric estimation of hurdle regression models with an application to {M}edicaid utilization.
\newblock {\em Journal of Applied Econometrics\/}~{\em 12\/}(3), 225--242.

\bibitem[\protect\citeauthoryear{Hopkins, Reeber, Forman, and Suermondt}{Hopkins et~al.}{1999}]{spambase_94}
Hopkins, M., E.~Reeber, G.~Forman, and J.~Suermondt (1999).
\newblock Spambase.
\newblock UCI Machine Learning Repository.
\newblock {DOI}: https://doi.org/10.24432/C53G6X.

\bibitem[\protect\citeauthoryear{Huber}{Huber}{2011}]{huber2011robust}
Huber, P.~J. (2011).
\newblock Robust statistics.
\newblock In {\em International encyclopedia of statistical science}, pp.\  1248--1251. Springer.

\bibitem[\protect\citeauthoryear{Hussain}{Hussain}{2018}]{studentperformancedata}
Hussain, S. (2018).
\newblock {Student Academics Performance}.
\newblock UCI Machine Learning Repository.
\newblock {DOI}: https://doi.org/10.24432/C50W30.

\bibitem[\protect\citeauthoryear{Kesgin, Kiraz, Kosunalp, and Stoycheva}{Kesgin et~al.}{2025}]{kesgin2025beyond}
Kesgin, K., S.~Kiraz, S.~Kosunalp, and B.~Stoycheva (2025).
\newblock Beyond performance: Explaining and ensuring fairness in student academic performance prediction with machine learning.
\newblock {\em Applied Sciences\/}~{\em 15\/}(15), 8409.

\bibitem[\protect\citeauthoryear{Kivaranovic and Leeb}{Kivaranovic and Leeb}{2021}]{kivaranovic2021length}
Kivaranovic, D. and H.~Leeb (2021).
\newblock On the length of post-model-selection confidence intervals conditional on polyhedral constraints.
\newblock {\em Journal of the American Statistical Association\/}~{\em 116\/}(534), 845--857.

\bibitem[\protect\citeauthoryear{Kleiber and Zeileis}{Kleiber and Zeileis}{2008}]{AERpackage}
Kleiber, C. and A.~Zeileis (2008).
\newblock {\em Applied Econometrics with {R}}.
\newblock New York: Springer-Verlag.

\bibitem[\protect\citeauthoryear{Kuchibhotla, Brown, Buja, Cai, George, and Zhao}{Kuchibhotla et~al.}{2020}]{kuchibhotla2020valid}
Kuchibhotla, A.~K., L.~D. Brown, A.~Buja, J.~Cai, E.~I. George, and L.~H. Zhao (2020).
\newblock {Valid post-selection inference in model-free linear regression}.
\newblock {\em The Annals of Statistics\/}~{\em 48\/}(5), 2953 -- 2981.

\bibitem[\protect\citeauthoryear{Le~Duy and Takeuchi}{Le~Duy and Takeuchi}{2021}]{le2021parametric}
Le~Duy, V.~N. and I.~Takeuchi (2021).
\newblock Parametric programming approach for more powerful and general lasso selective inference.
\newblock {\em International conference on artificial intelligence and statistics\/}, 901--909.

\bibitem[\protect\citeauthoryear{Lee, Sun, Sun, Taylor, et~al.}{Lee et~al.}{2016}]{lee2016exact}
Lee, J.~D., D.~L. Sun, Y.~Sun, J.~E. Taylor, et~al. (2016).
\newblock Exact post-selection inference, with application to the lasso.
\newblock {\em Annals of Statistics\/}~{\em 44\/}(3), 907--927.

\bibitem[\protect\citeauthoryear{Lee, Sun, and Taylor}{Lee et~al.}{2015}]{lee2015modelselection}
Lee, J.~D., Y.~Sun, and J.~E. Taylor (2015).
\newblock {On model selection consistency of regularized M-estimators}.
\newblock {\em Electronic Journal of Statistics\/}~{\em 9\/}(1), 608 -- 642.

\bibitem[\protect\citeauthoryear{McCullagh and Nelder}{McCullagh and Nelder}{1989}]{mccullagh1989generalized}
McCullagh, P. and J.~A. Nelder (1989).
\newblock {\em Generalized Linear Models}, Volume~37.
\newblock CRC Press.

\bibitem[\protect\citeauthoryear{Meinshausen, Meier, and B{\"u}hlmann}{Meinshausen et~al.}{2009}]{meinshausen2009p}
Meinshausen, N., L.~Meier, and P.~B{\"u}hlmann (2009).
\newblock P-values for high-dimensional regression.
\newblock {\em Journal of the American Statistical Association\/}~{\em 104\/}(488), 1671--1681.

\bibitem[\protect\citeauthoryear{Neufeld, Gao, and Witten}{Neufeld et~al.}{2022}]{neufeld2022tree}
Neufeld, A.~C., L.~L. Gao, and D.~M. Witten (2022).
\newblock Tree-values: selective inference for regression trees.
\newblock {\em The Journal of Machine Learning Research\/}~{\em 23\/}(1), 13759--13801.

\bibitem[\protect\citeauthoryear{Panigrahi and Taylor}{Panigrahi and Taylor}{2023}]{panigrahi2023approximate}
Panigrahi, S. and J.~Taylor (2023).
\newblock Approximate selective inference via maximum likelihood.
\newblock {\em Journal of the American Statistical Association\/}~{\em 118\/}(544), 2810--2820.

\bibitem[\protect\citeauthoryear{Pirenne and Claeskens}{Pirenne and Claeskens}{2024}]{pirenne2024parametric}
Pirenne, S. and G.~Claeskens (2024).
\newblock Parametric programming-based approximate selective inference for adaptive lasso, adaptive elastic net and group lasso.
\newblock {\em Journal of Statistical Computation and Simulation\/}~{\em 94\/}(11), 2412--2435.

\bibitem[\protect\citeauthoryear{Prerika, Nitika, and Kumar}{Prerika et~al.}{2025}]{Prerika}
Prerika, Nitika, and K.~Kumar (2025).
\newblock Email spam detection using artificial neural network with hybrid feature selection.
\newblock In L.~Garg, N.~Kesswani, and I.~Brigui (Eds.), {\em AI Technologies for Information Systems and Management Science}, Cham, pp.\  108--117. Springer Nature Switzerland.

\bibitem[\protect\citeauthoryear{Rasines and Young}{Rasines and Young}{2023}]{rasines2023splitting}
Rasines, D.~G. and G.~A. Young (2023).
\newblock Splitting strategies for post-selection inference.
\newblock {\em Biometrika\/}~{\em 110\/}(3), 597--614.

\bibitem[\protect\citeauthoryear{Rinaldo, Wasserman, G'Sell, et~al.}{Rinaldo et~al.}{2019}]{rinaldo2019bootstrapping}
Rinaldo, A., L.~Wasserman, M.~G'Sell, et~al. (2019).
\newblock Bootstrapping and sample splitting for high-dimensional, assumption-lean inference.
\newblock {\em Annals of Statistics\/}~{\em 47\/}(6), 3438--3469.

\bibitem[\protect\citeauthoryear{Shen, Gregory, and Huang}{Shen et~al.}{2024}]{shen2024post}
Shen, Q., K.~Gregory, and X.~Huang (2024).
\newblock Post-selection inference in regression models for group testing data.
\newblock {\em Biometrics\/}~{\em 80\/}(3), ujae101.

\bibitem[\protect\citeauthoryear{Taylor and Tibshirani}{Taylor and Tibshirani}{2018}]{taylor2018post}
Taylor, J. and R.~Tibshirani (2018).
\newblock Post-selection inference for-penalized likelihood models.
\newblock {\em Canadian Journal of Statistics\/}~{\em 46\/}(1), 41--61.

\bibitem[\protect\citeauthoryear{Tibshirani, Taylor, Lockhart, and Tibshirani}{Tibshirani et~al.}{2016}]{tibshirani2016exact}
Tibshirani, R.~J., J.~Taylor, R.~Lockhart, and R.~Tibshirani (2016).
\newblock Exact post-selection inference for sequential regression procedures.
\newblock {\em Journal of the American Statistical Association\/}~{\em 111\/}(514), 600--620.

\bibitem[\protect\citeauthoryear{Wainwright}{Wainwright}{2009}]{wainwright2009sharp}
Wainwright, M.~J. (2009).
\newblock Sharp thresholds for high-dimensional and noisy sparsity recovery using l1-constrained quadratic programming (lasso).
\newblock {\em IEEE transactions on information theory\/}~{\em 55\/}(5), 2183--2202.

\bibitem[\protect\citeauthoryear{Wasserman and Roeder}{Wasserman and Roeder}{2009}]{wasserman2009high}
Wasserman, L. and K.~Roeder (2009).
\newblock High dimensional variable selection.
\newblock {\em Annals of statistics\/}~{\em 37\/}(5A), 2178--2201.

\bibitem[\protect\citeauthoryear{White}{White}{1982}]{white1982maximum}
White, H. (1982).
\newblock Maximum likelihood estimation of misspecified models.
\newblock {\em Econometrica: Journal of the Econometric Society\/}~{\em 50}, 1--25.

\bibitem[\protect\citeauthoryear{Zhang and Cheng}{Zhang and Cheng}{2017}]{zhang2017simultaneous}
Zhang, X. and G.~Cheng (2017).
\newblock Simultaneous inference for high-dimensional linear models.
\newblock {\em Journal of the American Statistical Association\/}~{\em 112\/}(518), 757--768.

\bibitem[\protect\citeauthoryear{Zhao and Yu}{Zhao and Yu}{2006}]{zhao2006model}
Zhao, P. and B.~Yu (2006).
\newblock On model selection consistency of lasso.
\newblock {\em The Journal of Machine Learning Research\/}~{\em 7}, 2541--2563.

\bibitem[\protect\citeauthoryear{Zhao, Small, and Ertefaie}{Zhao et~al.}{2022}]{zhao2022selective}
Zhao, Q., D.~S. Small, and A.~Ertefaie (2022).
\newblock Selective inference for effect modification via the lasso.
\newblock {\em Journal of the Royal Statistical Society Series B: Statistical Methodology\/}~{\em 84\/}(2), 382--413.

\end{thebibliography}

\vskip .65cm
\noindent
University of South Carolina
\vskip 2pt
\noindent
E-mail: qshen@email.sc.edu
\vskip 2pt

\noindent
University of South Carolina
\vskip 2pt
\noindent
E-mail: gregorkb@stat.sc.edu

\noindent
University of South Carolina
\vskip 2pt
\noindent
E-mail: huang@stat.sc.edu


\end{document}